\documentclass[12pt]{article}
\input epsfig.sty
\newcommand{\be}[3]{\begin{equation}  \hlabel{#1#2#3}}     
\newcommand{\ee}{ \end{equation}}
\newcommand{\ba}{\begin{array}}
\newcommand{\ea}{\end{array}}
\newcommand{\NP}[3]{{\em Nucl. Phys.}{ \bf B#1#2#3}}

\renewcommand{\arraystretch}{1.8}

\newcommand{\hjump}[2]{{#2}}
\newcommand{\setref}[1]{ }
\newcommand{\href}[1]{{\ref{#1}}}
\newcommand{\hlabel}[1]{\label{#1} 
  }
\newcommand{\hcite}[1]
{\cite{#1}}
\newcommand{\hbibitem}[1]
{\bibitem{#1} }
\newcommand{\hepth}[7]{
{hep-th/#1#2#3#4#5#6#7}}

\begin{document}

\thispagestyle{empty}
\rightline{HUB-EP-96/58}
\rightline{hep-th/9611140}
\rightline{November 1996}
\vspace{1truecm}
\centerline{\bf \large Entropy of $N=2$ black holes and 
 their $M$-brane description}
\vspace{1.2truecm}
\centerline{\bf Klaus Behrndt\footnote{e-mail: 
 behrndt@qft2.physik.hu-berlin.de}
and Thomas Mohaupt\footnote{e-mail: mohaupt@qft2.physik.hu-berlin.de}}
\vspace{.5truecm}
\centerline{Humboldt-Universit\"at, Institut f\"ur Physik}
\centerline{Invalidenstra\ss e 110, 10115 Berlin}
\centerline{Germany}

\vspace{2.2truecm}


\vspace{.5truecm}

\begin{abstract}
In this paper we discuss the $M$-brane description for a $N=2$ black
hole.  This solution is a result of the compactification of
$M$-5-brane configurations over a Calabi-Yau threefold with arbitrary
intersection numbers $C_{ABC}$.  In analogy to the $D$-brane
description where one counts open string states we count here open
$M$-2-branes which end on the $M$-5-brane.
\end{abstract}

\bigskip \bigskip
\newpage

\noindent
{\bf \large 1. Introduction} \bigskip 

\noindent
It has been an open question for long time what kind of states are
associated with the Bekenstein--Hawking entropy of black holes.
In string theory we
came a major step closer in answering this question.  Many black holes
can be embedded into type II string theory as intersections of
$D$-branes. In this picture the horizon of the black hole becomes the
surface of these branes. This opens the possibility to identify the
microscopic states we are looking for with the open string states
ending on the $D$-branes. This way it was possible to give a
statistical interpretation for the Bekenstein-Hawking entropy of many
types of black holes. For compactifications of type II string theories
on $K3 \times T_2$ this has been done e.g.\ in \hcite{st/va} and the
results for $T_6$ compactifications are given in \hcite{ca/ma},
\hcite{ma/su}. A microscopic interpretation of the entropy for $N=2$
black holes of an orientifold compactification was given in
\hcite{ka/lo}. In this paper we are going to discuss a generic type II
Calabi--Yau compactification in the limit of large vector multiplet
moduli.  The Bekenstein--Hawking entropy for the corresponding 
black holes was found recently in \hcite{be/ca}.
Explicit solutions have been given in \hcite{be/ca}
for the double extreme limit and in \hcite{be} for
the general case.

\medskip

Before we start we will fix our notation (see \hcite{be/ca} and
refs. therein).  The $N=2$ supergravity includes one gravitational,
$n_{V}$ vector and $n_{H}$ hyper multiplets. In what follows we will
neglect the hyper multiplets, assuming that these fields are
constant. The bosonic $N=2$ action is given by 
\be010 
S \sim \int
d^{4}x \sqrt{G} \{ R - 2 g_{A\bar{B}} \partial z^A \, \partial
\bar{z}^B + {1 \over 4 } ( \Im {\cal N}_{IJ} F^I \cdot F^J + \Re {\cal
N}_{IJ} F^{I} \cdot {^{\star}F^{J}}) \} \,,
\ee 
where the gauge field part $F^I \cdot F^J \equiv F^I_{\mu\nu} F^{J \,
\mu \nu}$ and $I,J = 0,1,\ldots,n_V$.  
The complex scalar fields of the
vector multiplets $z^{A}$ ($A=1,\ldots,n_{V}$) 
parameterize a special
K\"ahler manifold with the metric $g_{A\bar{B}} = \partial_{A}
\partial_{\bar{B}} K(z,\bar{z})$, where $K(z,\bar{z})$ is the K\"ahler
potential.  Both, the gauge field couplings ${\cal N}_{IJ}$ and the
K\"ahler potential $K$ are given in terms of the holomorphic
prepotential $F(X)$ by
\be012 \ba{l} 
e^{-K} =
i (\bar{X}^I F_I - X^I \bar{F}_{I}) \ , \\ {\cal N}_{IJ} =
\bar{F}_{IJ} + 2i {(\Im F_{IL}) (\Im F_{MJ}) X^{L} X^{M} \over (\Im
F_{MN}) X^{M} X^{N}} 
\ea 
\ee 
with $F_{I} = {\partial F(X) \over \partial X^{I}}$ and $F_{MN} =
{\partial^{2} F(X)\over \partial X^{M} \partial X^{N}}$ (note that
these are not gauge field components).  The scalar fields $z^{A}$ are
defined by
\be014 
z^{A} = {X^{A} \over X^{0}} 
\ee 
and for the prepotential we take the cubic form 
\be016 
F(X) = {1 \over 6} {C_{ABC} X^{A} X^{B} X^{C} \over X^{0}} 
\ee 
with general constant coefficients $C_{ABC}$. In type II
compactification these are the classical intersection numbers of the
Calabi--Yau three--fold. 
When there exists a dual heterotic model, then
the cubic part of the prepotential contains both a classical piece,
which is linear in the heterotic dilaton field and quantum
corrections, that do not depend on the heterotic dilaton. In general
there will be further corrections consisting (in the type II picture)
of a constant part proportional to the Euler number of the Calabi Yau as
well as world--sheet instanton corrections which are exponentially
suppressed for large vector multiplet moduli. As explained in
\hcite{be/ca} these corrections are likewise supressed in the
Bekenstein--Hawking entropy formula for the black hole solutions that
we are going to consider here.

\medskip

The paper is organized as follows. 
In the \hjump{02}{next section} we
describe a 4-dimensional black holes solution and show the relation
to a 5-dimensional string, which is the intersection of three
$M$-5-branes. In \hjump{03}{section 3} we give a microscopic
interpretation of the entropy. Finally, we  \hjump{04}{summarize} our
results.

\bigskip \bigskip

\noindent 
{\bf \large 2. The black hole solution} \setref{02}
\bigskip \newline 
\noindent
In this section we give a black hole solution to the Lagrangian
(\href{010}). It is an axion-free solution, i.e.\ 
the scalar fields $z^A$ and as consequence also the couplings
${\cal N}_{IJ}$ are pure imaginary (adopting our conventions,
which were specified above). 
Moreover we will consider a solution that carries
$n_{V}+1$ of the possible $2(n_{V}+1)$ charges, which is not the most
general axion-free solution \hcite{be/ca}.  The general solution
(including axions and all eight charges possible for $n_{V}=3$) for
the case that only $C_{123}$ is non-trivial has been discussed in
\hcite{be/ka}.

\medskip

The solution we are going to analyze is given in terms of 
$n_V+1$ harmonic functions $H^A$ and
$H_0$ \hcite{be}
\be030
 \ba{l}
 ds^2 = - e^{-2 U} dt^2 + e^{2 U} d\vec{x} d\vec{x} \qquad , \qquad
 e^{2U} = \sqrt{H_0 \, {1 \over 6} C_{ABC} H^A H^B H^C} \\
 F^A_{mn} = \epsilon_{mnp}\partial_p H^A \quad , \quad 
 F_{0\; 0m} = \partial_m (H_0)^{-1} \quad , \quad  
 z^A = i H_0 H^A e^{-2U} 
\ea
\ee
(note that $F_{I \mu \nu} = {\cal N}_{IJ} F^{J}_{\mu \nu}$).
To be specific we choose for the harmonic functions
\be120
H^{A} = \sqrt{2}( h^{A} + {  p^{A} \over r} ) \qquad , \qquad 
H_{0} = \sqrt{2}( h_{0} + { q_{0} \over r } )
\ee
where $h^{A}$, $h_{0}$ are constant and related to the
scalar fields at infinity. The symplectic coordinates and the
Kahler potential are given by
\be121
 X^0 = e^U \quad , \quad 
 X^A = i \, H^A H_0 e^{-U} \quad , \quad
 e^{-K} = 8 (H_0)^2 e^{-2 U} \ .
\ee
The electric and magnetic charges are defined by integrals
over the gauge fields at spatial infinity
\be122
 \ba{l}
q_I = \int_{S^2_{\infty}} {\cal N}_{IJ} {^*F^J} =
  \int_{S^2_{\infty}} {\cal N}_{I0} {^*F^0}\,,  \\
p^I = \int_{S^2_{\infty}} F^J =
 \int_{S^2_{\infty}} F^A \ .
\ea
\ee
Thus, the black hole couples to $n_V$ magnetic gauge fields $F^{A}_{mn}$ and
one electric gauge field $F^0_{0m}$ (${\cal N}_{A0}=0$ for our solution). 
To get the mass we have to look on the asymptotic geometry. First, in 
order to have asymptotically a Minkowski space we have the constraint 
$ 4 h_{0}\, {1 \over 6} C_{ABC} h^{A} h^{B} h^{C} = 1$. Then
\be130
 e^{-2U} = 1 - {2 M \over r} \pm \cdots .
\ee
Hence the mass is given by
\be140
 M = {q_{0} \over 4 h_{0}}  + {1 \over 2} p^{A} h_{0} \, C_{ABC} h^{B} 
h^{C} \ .
\ee
Using (\href{121}) and calculating the central charge $|Z|$ we find that the
black hole, as expected, saturates the BPS bound
\be141
M^2 = |Z|^2_{\infty} = e^K_{\infty} (q_0 X^0 - p^A F_A)^2_{\infty}\,,
\ee
where the r.h.s.\ has to be calculated at spatial infinity 
($e^U_{\infty} =1$).

\medskip

On the other side if we approach the horizon all constants $h^{A}$ and
$h_{0}$ drop out. The area of the horizon depends only on the conserved
charges $q_0,p^A$. Furthermore, if $q_{0} C_{ABC} p^{A} p^{B} p^{C}>0$ the
solution behaves smoothly on the horizon and we find for the area and 
entropy $S_{BH}$
\be150
 A = 4 \, S_{BH} = 4 \pi \sqrt{4 q_{0} \, {1 \over 6} C_{ABC} p^{A} 
 p^{B} p^{C}}  \ .
\ee
When comparing to reference \hcite{be/ca} one has to take into account
that we have replace $F(X)$ by $-F(X)$ and $q_{0}$ by $-q_{0}$ in this
paper.  Note also that the $d_{ABC}$ used there is related to the
intersection numbers $C_{ABC}$ by $d_{ABC} = - \frac{1}{6} C_{ABC}$.

If the charges and parameters $h^{A}$ are positive then the area of
the horizon defines a lower bound for the mass.  Minimizing the mass
with respect to $h^A$ and $h_0$ gives us the area of the horizon
\hcite{fe/ka}
\be170
4 \pi M^{2}|_{min.} = A\,.
\ee
In this case 
\be160
h_{0} = {q_{0} \over c}  \qquad , \qquad h^{A} = {p^{A} \over c}
\ee
where $c^{4} = {2 \over 3}\, q_{0}C_{ABC} p^{A} p^{B} p^{C}$. For
these values all scalars are constant, i.e. coincides with their value
on the horizon ($z^{A} \equiv z^{A} |_{hor.}$). By this procedure we
get the double extreme black holes \hcite{ka/sh}. Taking this limit,
our solution (\href{030}) coincides with the solution given in
\hcite{be/ca}.

We have discussed only the cubic part (\href{016}) of the prepotential
and neglected the world--sheet instanton corrections.  This
approximation is justified as long as $|z^A| \gg 1$, which holds
whenever $H_0$ is large, which means the black hole decompactifies to
a string.

\medskip

There are many ways to get the solution (\href{030}) by compactification of
higher-dimensional configurations. On the type II side we have 
a Calabi--Yau
compactification, e.g.\ of three $D$-4-branes and a $D$-0-brane for type IIA
string theory. Alternatively we can see our solution as a compactification
of an intersection of three $M$-5-branes and a boost along the common
string.  Let us discuss the last possibility in more detail. If we have 
$n_{V} = 3$ and if only
$C_{123}$ is non--vanishing,
our solution (with 3 moduli $A=1,2,3$) corresponds to the
following intersection in 11 dimensions
\hcite{ts}
\be220
ds^2_{11} =  {1 \over (H^1 H^2 H^3)^{1 \over 3}} \left[ du dv + H_0 du^2
 + H^1 H^2 H^3 d\vec{x}^2 + H^A \omega_A \right] \ .
\ee
The case of identical harmonic functions has been discussed before in
\hcite{pa/to}.
This is a configuration where three $M$-5-branes intersect over a common string
and each pair of $M$-5-branes intersects over a 3-brane. In going to 4
dimensions we first compactify over $H^A \omega_A$, with $\omega_A$ defining
three 2-dimensional line elements. After this we are in 5 dimensions and
have a string solution with momentum modes parametrized by $H_0$ ($H^A$ are
parametrizing the $M$-5-branes). Generalizing this solution to a 
Calabi--Yau three--fold
with generic intersection numbers $C_{ABC}$
we find for this 5-dimensional string
solution
\be230
ds^2 =  {1 \over ({1 \over 6} C_{ABC} H^A H^B H^C)^{1 \over 3}} 
 \left( dv du  + H_0 du^2 + ({1 \over 6} C_{ABC}H^A H^B H^C ) d\vec{x}^2 
\right) \ .
\ee

Compactifying this string solution over $u$ yields our 4-dimensional
black hole solution (\href{030}). The electric gauge field results from
Kaluza-Klein reduction from 5 to 4 dimensions. 
In 5 dimensions we have only magnetic gauge
fields which are inherited by the  $D=4$ solution. In addition one of the
4-dimensional scalar fields is the compactification radius, which is
related to $|H_0|$ and thus $|H_0|\gg 1$ gives us the
decompactification limit, for which corrections due to the non--cubic
terms in the prepotential are small.

For the generic case ($C_{ABC} p^A p^B p^C \neq 0$) this
decompactified 5-dimensional string solution is non-singular and the
asymptotic geometry near the horizon is given by $AdS_3 \times
S_2$. Moreover every supersymmetric black hole (also the singular
once) can be understood as a compactification of a non-singular
configuration (e.g.\ self-dual 3- or 1-brane) with the asymptotic
geometry $AdS_{p} \times S_{q}$ \hcite{be/be}. These geometries play
an important role for supersymmetry restoration near the horizon (see
also \hcite{ka/gi}).

\bigskip

\bigskip \bigskip

\begin{figure}[t]
\begin{center}
\mbox{\epsfig{file=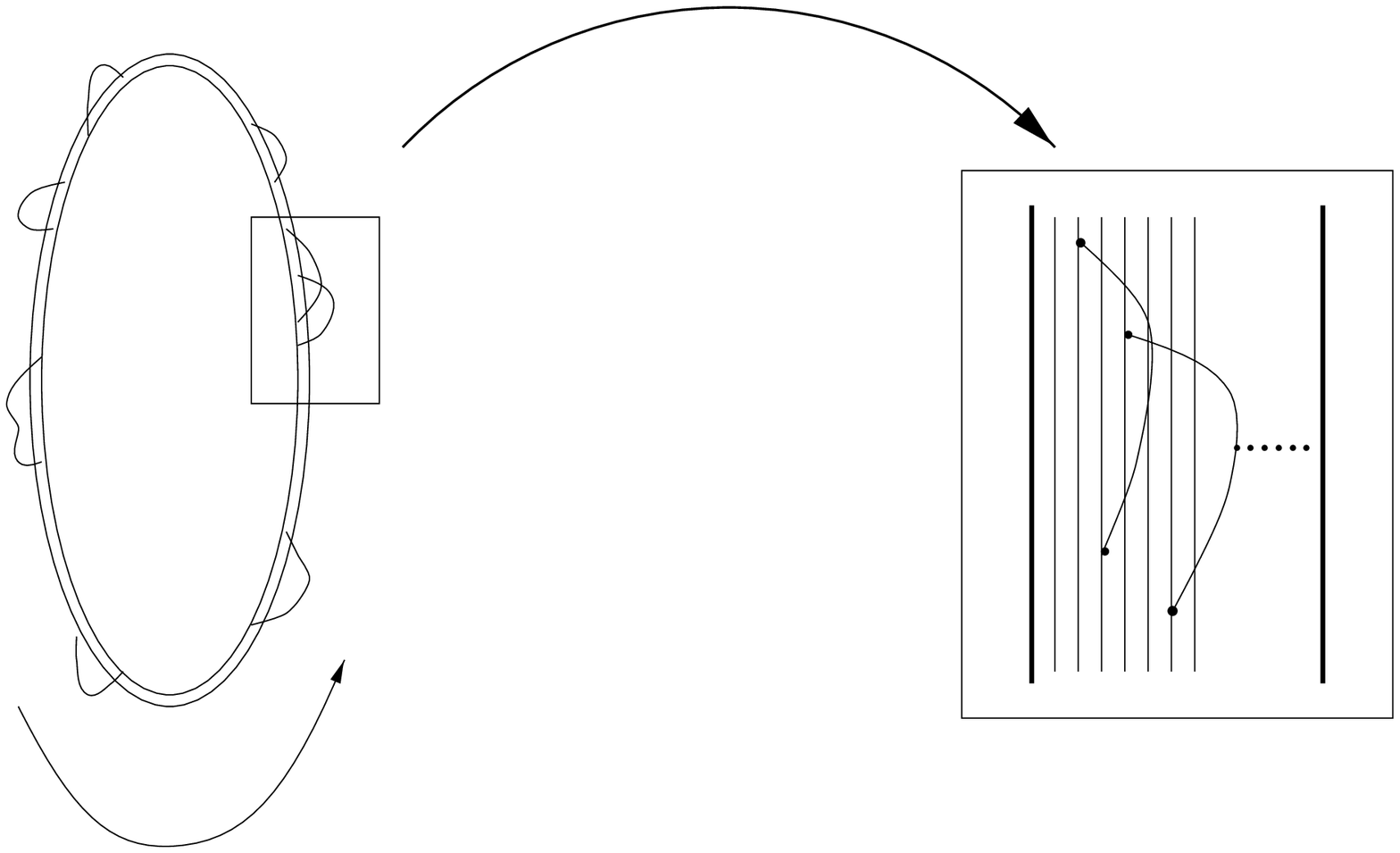, width=90mm}}
\end{center}
\noindent {\bf Fig.\ 1}\ \ {\it
This figure shows the 5-dimensional string with states
(momentum modes) travelling in one direction (BPS limit).
In the simplest case the string consists of $p^1 p^2 p^3$ layers.
}
\end{figure}

\noindent
{\bf \large 3. Microscopic interpretation of the entropy} \setref{03}
 \bigskip  \newline
\noindent
In this section we will propose a microscopic interpretation for the
entropy (\href{150}). Adapting the procedure of \hcite{st/va},
\hcite{ca/ma} we have to count the states of the 5-dimensional
magnetic string and to identify them as black hole states yielding the
Bekenstein-Hawking entropy. This procedure is known as $D$-brane
counting for the $N=4$ embedding where open string states are
counted which end on the $D$-brane\footnote{For an equivalent counting
of the NS-NS states see \hcite{ts2}.}. Here, however, the starting
point is 11-dimensional supergravity where we have 2-branes and
5-branes, which are called $M$-branes.
In analogy to type II open strings and 
$D$-branes, open $M$-2-branes can end on 
$M$-5-branes \hcite{st}. 
Therefore the states of the 5-dimensional string
are related to open membrane states attached to the 
$M$-5-branes. Since
one could in principle compactify to $D=10$ first, there is a direct
relation to open string states on $D$-4-branes.

A counting procedure for configurations of $M$-5-branes
with triple--intersections along strings has been proposed 
in \hcite{KlebTsey}.
We will adapt it to the case at hand. 
The first thing to notice is
that we are dealing with a BPS saturated solution (see (\href{141})),
which means that the momentum in the 5-dimensional 
string is purely 
left--moving (see figure 1). 
For a non--extreme solution, which does not satisfy
the BPS bound, there would be both left- and right--moving 
modes.
Since the 5-th dimension is compact, the momentum is quantized.
The integer quantum number $N_{L}$ has the interpretation of an
electric charge in four dimensions: $N_{L} = q_{0}$. 
The statistical entropy of left-moving states is given by
\be240
S_{stat} = \log d(N_L) = 2 \pi \sqrt{{1 \over 6} c_{eff} N_L}
\ ,
\ee
where $c_{eff}$ is the effective central charge associated to 
the 5-dimensional string.
Recall that in order to put momentum on the 5-dimensional 
string one has to
excite some internal degree of freedom. This is so because 
the spacelike
coordinate of the string world--sheet has been identified with 
the
5-th coordinate of the target space. Therefore translational 
invariance
(as well as transversality of physical states) forbids 
oscillating modes
in the 5-th direction. One can however excite some internal
degrees of freedom 
transversal to the string and let them move
in left direction around the string as described in figure 1.
These internal excitations have to be massless, because 
they are purely left--moving.
The 
entropy is found by counting in how many ways one can distribute
$q_{0}$ quanta of momentum among these massless internal modes.
One now
assumes that the massless internal degrees of freedom are 
described 
by a supersymmetric conformal sigma model with central charge
$c_{eff} = \frac{3}{2} D_{eff}$. The effective target space 
dimension
$D_{eff}$ is the number of massless excitations and for each 
dimension there is a bosonic and a fermionic world--sheet field,
which contribute $c=1$ and $c=\frac{1}{2}$ to the effective central
charge, respectively. The counting of states available with momentum
$q_{0}$ is equivalent to counting the number of left--moving oscillator 
states at level $N_{L} = q_{0}$
in a conformal field theory with central charge $c_{eff}$.
For large $q_{0}$, i.e. for $q_{0} \gg c_{eff}$ the corresponding
statistical entropy is given by (\href{240}).

The construction described so far is universal in the sense
that it applies to all 4--dimensional (and 5--dimensional) 
black holes with finite horizon that can be constructed
out of a 5--dimensional (6--dimensional) string. 
We now have to determine $c_{eff}$ by specifying 
the massless modes of the internal sector.
We will follow here the proposal made in \hcite{KlebTsey} for 
the counting of massless internal excitations of strings 
arising from
triple--intersections of $M$-5-branes. 
To explain this we take first some configurations of
$M$-5-branes which mutually intersect transversely
along 3-branes and have
triple--intersections along strings. Excitations of such a
configuration are described by open membranes which have 
boundaries on
different $M$-5-branes. In a generic situation no $M$-5-branes 
are sitting on
top of each other. Therefore the $M$-2-branes are stretched and there are
no massless states associated with them. There are however 
triple--intersections along the strings and a $M$-2-brane 
sitting at an
intersection describes the massless excitations of this string.  One
now has to assume that the relevant $M$-2-branes are those with three
boundaries, one sitting on each of the three intersecting $M$-5-branes.
Branes with less than 3 boundaries are giving a subleading
contribution, since the number of states grows with number of boundaries.
On the other side the number of branes with more than 3 boundaries
would grow too fast.

Next, a string inside a $M$-5-brane has four
transversal directions and therefore carries $c=6$. To get the
effective central charge one simply has to multiply this with the
number of intersections: $c_{{eff}} = 6 \cdot \sharp$
(intersections). In
order to construct a 4-dimensional black hole one now has 
to compactify from 11
to 5 dimensions while wrapping 4 dimensions of each $M$-5-brane 
in such a
way that all the strings resulting from triple--intersections 
are put
on top of each other. Then one wraps the resulting string 
around the 5-th dimension.

The authors of \hcite{KlebTsey} took a configuration of $p^{1}$
parallel $M$-5-branes which intersect with further 
$p^{2}$ parallel
$M$-5-branes in 3-branes. The number of intersections was $p^{1}p^{2}$ (see
figure 2). Including the third set of $p^{3}$ parallel 
$M$-5-branes
they got for the total number of intersections $p^{1}p^{2}p^{3}$. The
black hole was then constructed by toroidal compactification and
appropriate wrapping. The entropy of this configuration was
$S_{stat} = 2\pi
\sqrt{q_{0}p^{1}p^{2}p^{3}}$ which agreed with the 
Bekenstein--Hawking
entropy of the corresponding $D=4$ black hole solution
\hcite{KlebTsey}.

\begin{figure}[t]
\begin{center}
\mbox{\epsfig{file=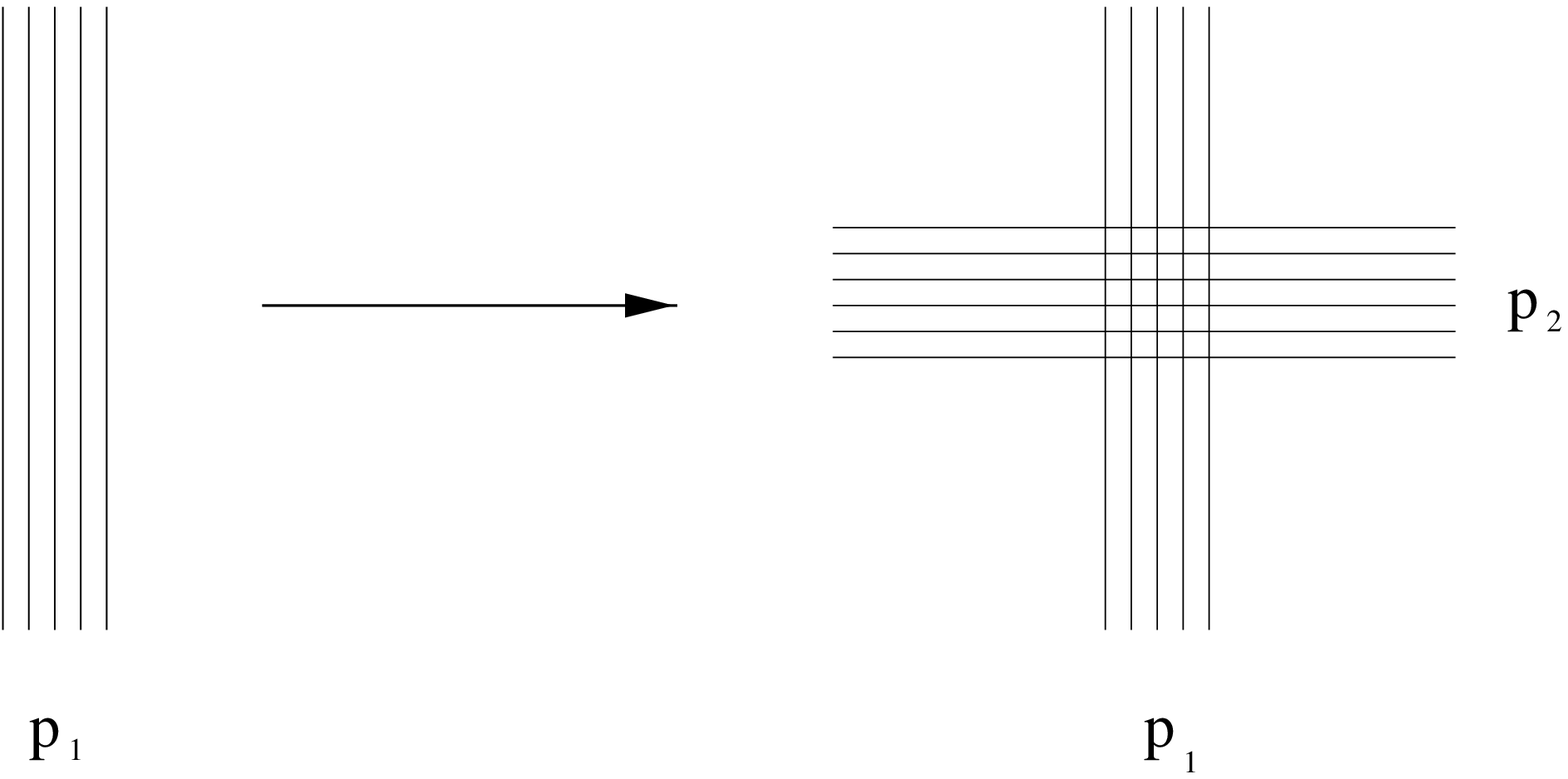, width=100mm}}
\end{center}
\noindent {\bf Fig.\ 2}\ \ {\it
Wrapping a brane $p^1$ times around a 4-cycle gives $p^1$ branes lying on top
of each other. An equivalent point of view is to assume that one has
$p^1$ parallel branes wrapping once around a 4-cycle.  
Intersecting two such 4-cycles yields $p^1 p^2$ branes
that lying on the common intersection.
}
\end{figure}

We will now generalize this to the case of a Calabi--Yau
compactification. Let us first consider the case where the only
non--vanishing intersection number of the Calabi--Yau is 
$C_{123}$. 
We first take $p^{1}$ $M$-5-branes and wrap them around $p^{1}$ 
homologous
but distinct primitive 4-cycles.  This generalizes the
case of parallel, but non--coinciding $M$-5-branes discussed 
before.
The reason why we insist on wrapping the branes around 
different 4-cycles is the same as in the flat space
situation discussed in \hcite{KlebTsey}: If $M$-5-branes
coincide, then the $M$-2-branes connecting them are no longer
stretched and can become massless. For $M$-2-branes with
three boundaries this would give a contribution 
$\sim (p^{1})^{3}$ to the entropy. But according the 
Bekenstein--Hawking formula (\href{150}) such contributions
should be associated with transversal triple--intersections
and not with coinciding $M$-branes. Especially there should
be no contribution if the self--intersection number 
$C_{111}$ is zero, which we assume here (the case $C_{111}
\not=0$ will be discussed below). 
Thus in order to interprete (\href{150})
one should take a configuration of non--coinciding
4-cycles that has triple--intersections in a discrete set
of points. 

Let us now proceed in giving an interpretation to
(\href{150}) for the case that only $C_{123}$ is 
non--vanishing.
We next wrap $p^{2}$ $M$-5-branes around cycles in the 
second and
$p^{3}$ $M$-5-branes around cycles in the third primitive 
homology class.  
Each triple of cycles with one cycle chosen from a different
homology classes intersects in $C_{123}$ points. 
The total number of intersections
is $C_{123}p^{1}p^{2}p^{3}$ (see also figure 2).
This can be written as $\frac{1}{6} C_{ABC} p^{A} p^{B} p^{C}$ using
that $C_{ABC}$ is symmetric. The resulting entropy is
\be241
S_{stat} = 
2 \pi \sqrt{q_{0}\, \frac{1}{6} C_{ABC} p^{A} p^{B} p^{C}}
\ee
which coincides with the Bekenstein--Hawking entropy found
in \hcite{be/ka} and recovered in (\href{150}). 

Let us now consider the case with generic $C_{ABC}$. A Calabi--Yau
manifold has $b_{4}= n_{V}$ different primitive 4-cycles which can
have various mutual as well as self--intersections.  The general
configuration is given by $p^{A}$ $M$-5-branes wrapped 
around distinct
4-cycles in the $A$-th primitive homology class, where
$A=1,\ldots,n_{V}$. The only point that deserves a further comment is
the case where self--intersections occur, i.e. if $C_{AAA}\not=0$ or
$C_{AAB} \not=0$. Let's consider the case $C_{111}\not=0$, which means that
three generic 4-cycles chosen from the first primitive homology class
triple-intersect in $C_{111}$ points.
Then the total number of intersection points is
$\frac{1}{6} C_{111} p^1 (p^1 - 1) (p^1 -2)$. The factor of 
$\frac{1}{6} = \frac{1}{3!}$
is needed to avoid overcounting, because now all the branes
are in the same homology class.
(When counting intersections between branes in different
classes, a factor $\frac{1}{6}$ was introduced for different
reasons, namely to compensate for permutations arising from
the summation over the indices. )
Moreover intersections occur only between different 
cycles so that we have to replace 
$(p^1)^3$ by $p^1 (p^1 -1) (p^1 -2)$.
Note that this number is divisible by 6.
In the limit of large charges the number of intersections
is dominated by the cubic piece and we get $\frac{1}{6} C_{111} (p^1)^3$. 
%
%
The discussion of non--vanishing intersection numbers $C_{AAB}$ is
similar and one finds $\frac{1}{6} C_{AAB} p^A(p^A-1) p^B \sim
\frac{1}{6} (p^A)^2 p^B$ intersection
points. As a result the microscopic entropy formula for a Calabi--Yau
compactification with generic intersection numbers is given by
(\href{241}) and coincides with the Bekenstein--Hawking 
entropy (\href{150}).

\medskip

\bigskip \bigskip

\noindent
{\bf \large 4. Discussion} \setref{04} \bigskip 

\noindent
In this paper we have given a microscopic interpretation for the
Bekenstein-Hawking entropy of a $N=2$ black hole. Our solution has been
obtained by a $M$-theory compactification over a Calabi--Yau 
three--fold with
arbitrary self--intersection numbers $C_{ABC}$. In 11 dimensions this
solution corresponds to a configuration of $M$-5-branes with 
triple--intersections along strings. We have
counted states on the $M$-5-branes, however not open string states as in the
$D$-brane technique but open membrane states. Following \hcite{KlebTsey}
we assumed that the leading contribution to the entropy comes 
from massless open $M$-2-branes with three boundaries, 
each sitting on a different $M$-5-brane. Since such $M$-2-branes can
only become massless in the vicinity of a triple--intersection
of $M$-5-branes we considered a configuration of $M$-5-branes
with triple--intersections along strings.  
Compactifying this
configuration over the Calabi--Yau threefold, 
i.e.\ wrapping the $M$-5-branes around
4-cycles, we got in 5 dimensions a magnetic strings with momentum modes
corresponding to the open membrane states. The magnetic charges counts how
many times we had wrapped a $M$-5-brane around a 4-cycle.
Or, from a different point of view, the magnetic charge counts the
number of parallel $M$-5-branes that are wrapped once around the 4-cycle.
The electric charge gives then the number of momentum modes.
The intersection pattern of the $M$-5-branes was governed by the
intersection form $C_{ABC}$ of the Calabi--Yau. The only freedom
consisted off choosing the numbers $p^A$ of $M$-5-branes that we wrapped 
around 4-cylces in a given primitive homology class. The number of
intersection points was found to be $\frac{1}{6}C_{ABC} p^A p^B p^C$,
which is precisely the number one needs to reproduce the
Bekenstein--Hawking entropy by state counting. We also note here
that the structure of the entropy formula (\href{241}) is very similar
to the one of the 5--dimensional black hole discussed by 
Strominger and Vafa\hcite{st/va}. Both solutions are related
by replacing $D$-branes by $M$-branes, the $K3$ surface
by a Calabi--Yau space and internal 2-cycles by 4-cycles. 
It would be very interesting to analyse these parallels
in more detail. This could help to devellop the world--volume
theory of curved $M$--branes. In analogy to the findings
of \hcite{st/va} one expects this to be a partially 
topologicially twisted theory, which should encode geometrical
properties of the wrapping cycles inside the Calabi--Yau space.

\medskip

The inclusion of all possible self-intersections of the 4-cycles
reveals some new features. For many Calabi--Yau 
compactification we do not need
anymore four independent charges to have a black hole with a
non-singular horizon. A simple cubic term, like $C_{333}$, in the
prepotential makes already the horizon non-singular. In the discussion
of black holes that are non-singular only due to self-intersections of
4-cycles we have, however, to keep in mind that this procedure can only be
trusted as long as all charges are large, which is related to a large
black hole in 4 dimensions. If the charges become small (or even
vanish) there are corrections to be taken into account, which can be
parametrized by additional terms to the prepotential (\href{016}). 

Throughout the paper we have restricted ourselves to the cubic
part of the prepotential, which for type II A strings is valid in the 
limit of large K\"ahler moduli. Moreover the radius of the 5-th
direction has to be taken even larger then the size of the Calabi--Yau
manifold. In terms of charges this means $q_{0} \gg c_{eff} \sim
C_{ABC}p^{A}p^{B}p^{C}$, i.e.\ that we just began
to explore a large moduli space starting from a particular corner.
Since the type II dilaton sits in a hyper multiplet, the prepotential
is not expected to get perturbative or non--perturbative quantum
corrections. There are however $\alpha'$-correction, both perturbative
and non--perturbative. The perturbative corrections yield a constant
term proportional to the Euler number of the Calabi--Yau, whereas
the non--perturbative corrections result in an infinite series 
of world--sheet instantons. One therefore has to expect that the
entropy depends on the Euler number as well as on the world--sheet
instanton numbers that count the rational holomorphic curves
inside the Calabi--Yau. 
Indeed, as discussed in \hcite{be/ca} the constant term enters
the Bekenstein--Hawking entropy for some type of black hole 
solution. But this term comes with a trancendental prefactor
$\zeta(3)$ which is hard to reproduce by state counting\footnote{
T.M. thanks Bernard de Wit for pointing this out to him.}. 
Therefore
further contributions to the entropy from world--sheet instantons
are needed in order make a statistic interpretation possible.
Note that it is natural to include the constant term in the
world--sheet instanton series as its zero mode.
To see this  recall that the correction to the cubic term of the
prepotential has the form \hcite{CY}
\[ 
- \frac{\chi}{2} \zeta(3)
+ \sum_{m_i} n_{m_i} Li_{3}(e^{-m_{i}t_{i}}) \,,
\]
where $\chi$ is the Euler number, 
$m_{i}\not=0$ is a multi--index that  
labels world--sheet instantons, $n_{m_i}$ is the number of 
world--sheet instantons of type $m_i$ and $t_{i}$ are the moduli.
Since  $\zeta(3) = Li_{3}(1)$ one can include the constant
term in the sum by defining $n_0 = - \frac{\chi}{2}$. 

We would also like to mention that there is a class of $N=2$ string
models, where the cubic part of the prepotential is exact. These
models are very similar to $N=4$ models in that there are no
perturbative quantum corrections. (We are using here the heterotic
string picture, where the world--sheet instanton corrections of the
type II picture are mapped to quantum loop corrections.) This class
contains the II A orientifold, for which the microscopic entropy was 
studied in \hcite{ka/lo}. In \hcite{FHSV} the Calabi--Yau
threefold corresponding to this model was described.  It
is a self--mirror, implying that the Euler number as well as all
world--sheet instanton corrections must vanish.  This model might be
good laboratory for a deeper study of the geometric structure behind
the $M$-brane picture discussed here.  Note also that the vanishing of
the constant and of the instanton correction to the prepotential are
enforced in once.  This supports our speculation above about the close
relationship of these terms.

Finally we would like to recall that our black hole solution
was based on the prepotential, which only takes into account
the minimal terms in the effective action, i.e. those
with the minimal number of derivatives. Since it is
well known that string--effective action contain an infinite
series of higher derivative terms, it is interesting to ask
how these will modify the picture of stringy black holes
that we have today.

\vspace{1.5truecm}

\newpage

\noindent
{\bf Acknowledgements}  \medskip \newline
The work is  supported by the DFG. We would like to thank 
Dieter L\"ust and Edward Derrick for many useful discussions and
comments.

\renewcommand{\arraystretch}{1}



\begin{thebibliography}{aaa}

\hbibitem{st/va}
A. Strominger and C. Vafa, {\it Microscopic origin of the
Beckenstein-Hawking entropy}, {\tt \hepth9601029};
R. Dijkgraaf, E. Verlinde and H. Verlinde, {\it Counting dyons in
$N=4$ string theory}, {\tt \hepth9607026}.
\hbibitem{ca/ma}
C.G. Callan and J.M. Maldacena, {\it $D$-brane approach to black hole
quantum mechanics}, {\tt \hepth9602043};
J.M. Maldacena and A. Strominger, {\it Statistical entropy of four-dimensional
extreme black holes}, {\tt \hepth9603060}.
\hbibitem{ma/su}
J.M. Maldacena and L. Susskind, {\it $D$-branes and fat black holes},
{\tt \hepth9604042}.
\hbibitem{ka/lo}
D.M. Kaplan, D.A. Lowe, J.M. Maldacena and A. Strominger,
{\it Microscopic entropy of $N=2$ extremal black holes}, {\tt 
 \hepth9609204}.
\hbibitem{be/ca}
K. Behrndt, G. Lopes Cardoso, B. de Wit, R. Kallosh, D. L\"ust and
T. Mohaupt, {\it Classical and quantum $N=2$ supersymmetric black holes},
{\tt \hepth9610105}.
\bibitem{be}
K. Behrndt, {\it Quantum corrections for $N=4$ black holes and $D=5$
strings}, {\tt \hepth9610232}.
\hbibitem{be/ka}
K. Behrndt, R. Kallosh, J. Rahmfeld, M. Shmakova, W.K. Wong, {\it
STU black holes and string triality}, {\tt \hepth9608059}; \\
G. Lopes Cardoso, D. L\"ust and T. Mohaupt, {\it Modular Symmetries of N=2
  Black Holes}, {\tt \hepth9608099}.
\hbibitem{ka/sh}
R. Kallosh, M. Shmakova and W.K. Wong, {\it Freezing of moduli by
 $N=2$ dyons}, {\tt \hepth9607077}.
\hbibitem{fe/ka}
S. Ferrara, R. Kallosh and A. Strominger, {\it $N=2$ extremal black holes},
{\tt \hepth9508072}, \\
R. Kallosh and S. Ferrara, {\it Supersymmetry and attractors} , 
{\tt \hepth9602136}.
\hbibitem{ca/os}
P. Candelas, X.C. de la Ossa, P.S. Green and L. Parkes, {\it
A Pair of Calabi-Yau manifolds as an exactly soluble superconformal theory},
\NP359(1991)21.
\hbibitem{ts}
A.A. Tseytlin, {\it Harmonic superposition of $M$-branes}, 
{\tt \hepth9604035}.
\hbibitem{pa/to}
G. Papadopoulos and P.K. Townsend, {\it Intersecting M-branes},
{\tt \hepth9603087}.
\hbibitem{be/be}
K. Behrndt and E. Bergshoeff, {\it A note on intersecting $D$-branes
and black hole entropy}, {\tt \hepth9605216}.
\hbibitem{ka/gi}
A. Chamseddine,  S. Farrara, G.W. Gibbons and R. Kallosh, {\it Enhancement
of supersymmetry near 5-D black hole horizon}, {\tt \hepth9610155 }.
\hbibitem{ts2}
F. Larsen and F. Wilczek, {\it Internal structure of black holes},
{\tt \hepth9511064};
A.A. Tseytlin, {\it Extreme black hole entropy from conformal
string sigma model}, {\tt \hepth9605091}.
\hbibitem{st}
A. Strominger, {\it Open $p$-branes}, {\tt \hepth9512059};\\
P.K. Townsend, {\it $D$-branes form $M$-branes}, {\tt \hepth9512062}.
\hbibitem{KlebTsey} I. Klebanov and A. Tseytlin, {\it Intersecting M-branes
as Four--Dimensional Black Holes}, {\tt \hepth9604166}.
\hbibitem{FHSV} S. Ferrara, J. A. Harvey, A. Strominger and
C. Vafa, {\it Second--Quantized Mirror Symmetry}, {\em Phys. Lett.}
{\bf B 361} (1995) 59, {\tt \hepth9505162}.
\hbibitem{CY} P Candelas, X. de la Ossa, P. Green and L. Parkes,
{\it A pair of Calabi--Yau manifolds as an exactly soluble
superconformal theory}, {\em Nucl. Phys.} {\bf B 359}
(1991) 21, \\
S. Hosono, A. Klemm, S. Theisen and S.-T. Yau,
{\it Mirror symmetry, mirror map and Applications to complete
intersection Calabi--Yau spaces}, 
{\em Nucl. Phys.} {\bf B 433} (1995) 501.
\end{thebibliography}
\end{document}